\documentclass[letter]{aa} 
\usepackage{graphicx}
\usepackage{txfonts}
\usepackage{natbib}
\bibpunct{(}{)}{;}{a}{}{,} 

\newcommand{\Kalpha}{K$_{\alpha}$}

\newcommand{\nden}[1]{$n_{\rm {#1}}$}

\newcommand{\be}{\begin{displaymath}}
\newcommand{\ee}{\end{displaymath}}
\begin{document}
  \title{\emph{XMM-Newton} RGS spectrum of  RX\, J0720.4$-$3125: \\
An absorption feature at 0.57~keV\thanks{Based on observations obtained with
 {\em XMM-Newton}, an ESA science mission with instruments and contributions directly funded by ESA Member States and NASA} }

   \author{V. Hambaryan
          \inst{1}
          \and
	  R. Neuh\"auser
          \inst{1}
          \and
	  F. Haberl
          \inst{2}
	  \and
	  M.M. Hohle
          \inst{1,2}
         \and
	  A.D. Schwope
          \inst{3}
          }

   \institute{Astrophysikalisches Institut und Universit\"ats-Sternwarte, Universit\"at Jena, Schillerg\"a\ss chen 2-3, 07745 Jena, Germany\\
              \email{vvh@astro.uni-jena.de}
         \and
             Max-Planck-Institut f\"ur extraterrestrische Physik, Giessenbachstrasse, D--85741 Garching, Germany
	\and
		Astrophysikalisches Institut Potsdam, An der Sternwarte 16, D-14482 Potsdam, Germany
             }

   \date{Received ; accepted }

 
  \abstract
{}
   {By measuring the gravitational redshift of spectral features in the spectrum of thermal radiation 
    emitted from neutron stars, useful constraints for the equation of state of superdense matter
    can be obtained via an estimate of the mass-to-radius ratio. 
    We searched for spectral line features in the high-resolution X-ray spectrum 
    of the isolated neutron star RX\, J0720.4$-$3125.}
   {Our target was observed by \emph{XMM-Newton} on many occasions. 
    We used the \emph{XMM-Newton~SAS} task \emph{rgscombine} in order to create a
    co-added RGS spectrum 
    of  RX\, J0720.4$-$3125. We modeled the resulting spectrum with absorbed blackbody radiation with 
     a Gaussian absorption features using the XSPEC package.}
   {We found a narrow absorption feature at 0.57~keV in the co-added RGS spectrum of the isolated neutron star 
    RX\, J0720.4$-$3125 with an equivalent width of $1.35\pm0.3$~eV and FWHM $\sim 6.0$~eV. The feature was 
    identified with an absorption line of highly ionized oxygen \ion{O}{VII},  most probably originating 
    in the ambient medium of RX\, J0720.4$-$3125. 
    An extensive investigation with the photo-ionization code CLOUDY indicates the possibility that the 
    optical flux excess observed in the spectrum of  RX\, J0720.4$-$3125 at least partially originates 
    in a relatively dense (e.g. $\mathrm{n}_{H} \sim 10^8 \mathrm{cm}^{-3}$) slab, 
    located in the vicinity of the neutron star (e.g. $\sim \mathrm{10}^{10} \mathrm{cm}$).}
   {}


   \maketitle
%

\section{Introduction}

   The study of  thermally emitting and radio-quiet, nearby, isolated neutron stars 
(INSs) may allow an important input to our understanding of neutron stars. 
In particular, measuring the gravitational redshift of a spectral feature  
in the spectrum of thermal radiation emitted from the neutron star surface (atmosphere) 
may provide a useful constraint for various equations of state 
for  superdense matter. Independent of the estimate of the neutron star 
radius \citep[e.g.][]{2004NuPhS.132..560T} 
from the thermal spectrum of an INS, it will also allow us to
directly estimate the mass-to-radius ratio.  
RX\, J0720.4$-$3125, originally discovered as a soft X-ray source during the
\emph{ROSAT} All-Sky Survey by \cite{1997A&A...326..662H}, is a special case among the INS.
Clear spectral variations on times scales of years were detected during {\em XMM-Newton} 
observations in the data of the high spectral resolution Reflection Grating Spectrometers (RGS) 
\citep{2004A&A...415L..31D} and the imaging-spectroscopic instrument EPIC~pn
\citep{2006A&A...451L..17H}. 
However, during individual RGS observations 
 the number of counts received from  RX\, J0720.4$-$3125 in a single 
resolution element were not enough to study the spectrum in detail. 

From this point of view, it is interesting to study a co-added {\em XMM-Newton} RGS spectrum
of RX\, J0720.4$-$3125, despite of the spectral variations shown by the INS.

\section{Observations and data reduction}

RX\, J0720.4$-$3125 has been observed many times by \emph{XMM-Newton} 
\citep{2008A&A_sub1} and we focus here on the data collected with RGS 
\citep{2001A&A...365L...7D} from the 14 publicly available 
observations, in total presenting about 500~ks of effective exposure time.

\begin{table}
\caption[]{\emph{XMM-Newton} RGS1/2 observations of  RX\, J0720.4$-$3125}
\label{obslog}      
\centering                          
\begin{tabular}{cccc}
\hline\noalign{\smallskip}
Obs. ID & Obs. Date Start & Exposure & Effective exposure \\
        &   JD - 2400000 &  ksec    &    ksec            \\
\hline
0124100101 & 51678.1503935 &  65.87   & 15.45 \\
0132520301 & 51870.2714931 &  30.91   & 30.24 \\
0156960201 & 52585.2443171 &  30.24   & 29.60 \\
0156960401 & 52587.3090509 &  32.04   & 31.03 \\
0158360201 & 52762.0499190 &  81.63   & 73.43 \\
0161960201 & 52940.2405093 &  44.92   & 44.67 \\
0164560501 & 53147.9273380 &  51.95   & 44.36 \\
0300520201 & 53488.8618634 &  53.31   & 50.69 \\
0300520301 & 53636.4891551 &  53.01   & 51.38 \\
0311590101 & 53687.4349306 &  39.71   & 38.91 \\
0400140301 & 53877.6977662 &  21.91   & 21.57 \\
0400140401 & 54044.9718634 &  21.91   & 21.69 \\
0502710201 & 54226.2093171 &  21.91   & 21.57 \\
0502710301 & 54421.7184259 &  24.92   & 24.59 \\
\hline                                   
\end{tabular}
\end{table}
%

The data were reduced using  standard threads from the 
{\em XMM-Newton}~data analysis package SAS version 8.0.1. 
We reprocessed all publicly available data (see Table~\ref{obslog}) with the standard metatask \emph{rgsproc} 
with the aim to co-add the spectra with the task \emph{rgscombine}. To determine good time intervals free of 
background flares, we used the filtering expression `rate~$<$~1.0' on the background light curves.
This reduced the total exposure time  by $\sim$~13\%. First order source and 
background spectra were produced from the 
cleaned events. A model background spectrum was also generated with the task {\it rgsbkgmodel} that also 
was used for the background correction as an additional check for possible phenomena related 
to background variations seen during some of the observations. 
No significant differences between RGS spectra reduced in these two different ways were seen. 

The background subtracted co-added spectra of RX\, J0720.4$-$3125 revealed the existence of a relatively
narrow absorption-like feature at $\sim$0.57~keV. No sharp variation is present in the spectral response 
of the instrument, indicating an astronomical origin. A similar narrow absorption feature was reported in the 
{\em XMM-Newton} RGS spectrum of another INS  RX\, J1605.3+3249 at 0.576~keV 
by \cite{2004ApJ...608..432V}.

\begin{figure}
   \centering
   \includegraphics[width=8.6cm]{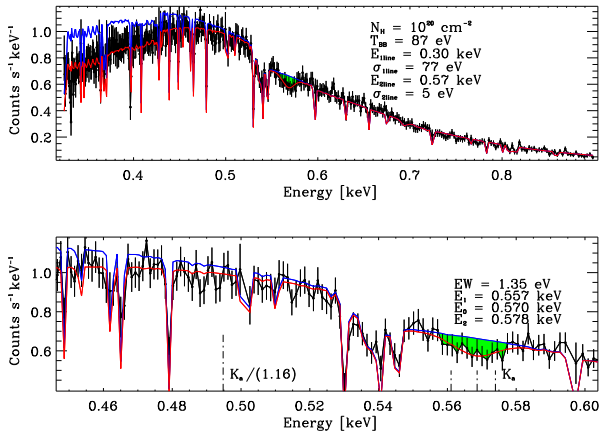}
      \caption{{\em XMM-Newton} high-resolution co-added spectrum of the isolated neutron star RX\, J0720.4$-$3125.
	An absorbed blackbody (blue) and an absorbed blackbody with two additional Gaussian
	 absorption lines (red, a wide absorption feature at 0.3 keV and a narrow feature at 0.57~keV) are shown. 
        The equivalent width and the FWHM of the narrow absorption line (green), which we associate with OVII, 
          were determined at 1.35~eV and $\sim$6.0~eV, respectively.
	  Different spectral line positions at gravitational redshift ($g_r = 0$) are indicated.
              }
         \label{FigSpec}
   \end{figure}

We fitted the resulting co-added RGS1\footnote{The RGS2 spectrum was used only for consistency checks; 
in the energy range of 0.5-0.59~keV it has a significantly reduced effective area owing to the 
failure of CCD4 in September 2000, see {\em XMM-Newton} 
Users Handbook} spectrum of RX\, J0720.4$-$3125   (see Fig.~\ref{FigSpec}) using
the \emph{XSPEC} package version~12.5.0. We fitted an absorbed blackbody model 
with a broad (0.08~keV) Gaussian absorption feature at 0.3~keV, well known from 
{\em XMM-Newton} EPIC~pn spectral studies \citep{2007Ap&SS.308..181H}.

Another additive Gaussian absorption component was used to fit the feature at 0.57~keV.
We used this model to estimate the equivalent width of the detected absorption feature at 0.57~keV.
As the continuum level the absorbed blackbody model including the broad Gaussian line in absorption
 was used. With different levels (normalizations) of continuum and absorption lines an error of the 
equivalent width was determined. Results of the fit and equivalent width estimates are presented in Fig.~\ref{FigSpec}.

\begin{figure}
   \centering
   \includegraphics[width=8.0cm,clip=]{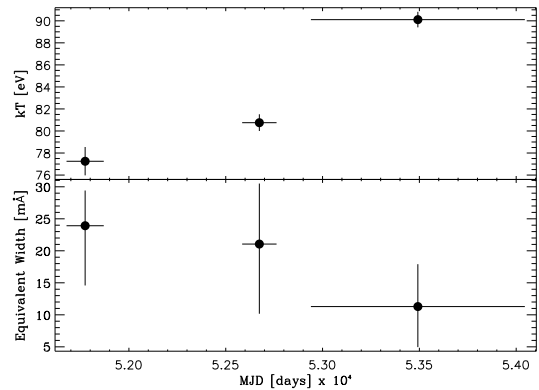}
      \caption{Equivalent width of the absorption feature at 0.57~keV and 
       blackbody temperature derived from the RGS spectra of RX\, J0720.4$-$3125 from three different epochs.
              }
         \label{FigeqwTr}
   \end{figure}

As already mentioned, RX\, J0720.4$-$3125 showed spectral variations, 
first  noticed from the  comparison of RGS spectra observed at different epochs \citep{2004A&A...415L..31D}. 
 After the detection of the absorption feature at 0.57~keV in the co-added RGS spectrum 
of RX\, J0720.4$-$3125, we examined
 its behavior over time. We divided all publicly available observations into three groups 
from different epochs. In the first group, we combined the RGS spectra  
of RX\, J0720.4$-$3125 from the year 2000 (ObsIds 0124100101 and 0132520301), for the second group 
the spectra between Nov. 2002 and May 2003 (0156960201, 0156960401 and 0158360201,  
and in the third group, the spectra  from Oct. 2003 to Nov. 2007.
This devision was inferred from our recent study on the nature of global variations observed 
 from RX\, J0720.4$-$3125 \citep{2008A&A_sub1} which was mainly based on the analysis of the 
EPIC~pn data. In this way the variations in the blackbody temperature and the depth of the broad absorption feature in 
the spectra within each group of observations \citep[see, Fig.~7,][]{2008A&A_sub1} are kept to a minimum and below
the sensitivity of the RGS instrument. The results of this analysis are presented in Fig.\ref{FigeqwTr}. 
It shows a tendency for a reciprocal dependence of the measured equivalent width on the derived
 blackbody temperature. To decide whether this is due to 
 pure flux (blackbody temperature) variations or to changes in the number of atoms
  in the line of sight, we fitted the three data sets 
with linked parameters of absorption features and the blackbody temperature was left free. 
The estimated equivalent widths are consistent with the variations when parameters of the
narrow absorption feature were left free (Fig.~\ref{FigeqwTr}). However, this 
needs confirmation with higher S/N ratio data sets. 

\section{Discussion}

We report the detection of an absorption line at 0.57~keV in the {\emph XMM-Newton} RGS
spectra of RX\, J0720.4$-$3125. Since there is no uncalibrated instrumental feature in the 
energy range of 0.55$-$059~keV (21$-$22.5\AA), this line is clearly of astronomical origin.
Whether it originates  at the surface (atmosphere) of 
 the neutron star, in its vicinity or in the interstellar medium
is not clear a priori.

According to the atomic database ATOMDB\footnote{http://cxc.harvard.edu/atomdb}, 
 a few relatively strong spectral lines are expected 
in the energy range of  0.56$-$1.0~keV 
 (with a typical value of gravitional redshift of INS $\sim 0.2$)
 from plasma with a temperature of $\sim 10^{6}$~K, 
mainly belonging to highly ionized oxygen (\ion{O}{VII}, \ion{O}{VIII}) and neon (\ion{Ne}{IX}).
Among them the \ion{O}{VII} resonance \Kalpha~ spectral line at 0.574~keV 
has a maximal intensity and 
is compatible with our detection without any significant redshift.
It is likely that the absorption feature consists of a number of blended absorption lines. 
However, given the data quality (i.e., spectral resolution and S/N ratio) it is difficult to identify all of them. 

Moreover, similar absorption  features were detected in high-resolution X-ray spectra of 
other galactic and extragalactic sources 
({\em but not in all cases}\footnote{\cite{2004ApJ...605..793F}  did not detect the absorption line of \ion{O}{VII} 
in the spectrum of Cyg X-2. According to \cite{2005ApJ...624..751Y}, 3 of 10 Galactic low-mass X-ray binaries show absorption of 
 \ion{O}{VII} and/or  \ion{O}{VIII}. } 
see, eg., \citet{2004ApJ...605..793F}, \citet{2005ApJ...629..700N}, \citet{2005ApJ...624..751Y},  \citet{2006ApJ...644..174F}, \citet{2007ApJ...669..990B}).
This, as well as its relatively narrow nature \citep{2006ApJ...636L.117C, 1997ApJ...476L..47P}, 
make it more plausible that it originates in the  circumstellar/interstellar medium along the line of sight towards RX\, J0720.4$-$3125.
Indeed, the estimates of strength of the magnetic field of B$_{dp}\sim 2.4\times 10^{13}$~G (from dipole braking) 
and B$_{cyc} \sim 5.6\times ^{13}$~G (proton-cyclotron absorption feature) of RX\, J0720.4$-$3125
predict 20--60eV broadening of spectral features originating in the 
magnetized atmosphere of the INS \citep{1997ApJ...476L..47P}.
Nevertheless, with a gravitational redshift of INS of $\sim$~0.16, a strong lines 
of \ion{O}{VIII} at 0.653~keV are expected at 0.56~keV. In this case the \ion{O}{VII} 
resonance \Kalpha~ spectral line from the INS atmosphere is expected at 0.49 keV. 
A very faint absorption feature is present in 
the spectrum of  RX\, J0720.4$-$3125(Fig.~\ref{FigSpec}). However, given the fitted blackbody temperature ($\sim$~1MK) and 
the magnetic field strength (significantly increasing the binding energy of atoms) of  RX\, J0720.4$-$3125, 
it is quite difficult to reconcile the presence of significant number of \ion{O}{VIII} or 
other ions (e.g., \ion{Ne}{IX}, with higher excitation temperature) in the atmosphere of 
RX\, J0720.4$-$3125.

If the detected absorption line has an origin in the circumstellar/interstellar medium, it
may be identified with the highly ionized oxygen \ion{O}{VII}. 
Indeed, according to the atomic database (ATOMDB) the strongest 
laboratory-measured spectral lines  from \ion{O}{VII} in the energy
range 0.55$-$0.59~keV are:
0.5611, 0.5686 and 0.5740~keV.
However, the first two spectral lines are forbidden and only in specific conditions (density $ \sim$10$^{10}$~cm$^{-3}$, 
as well as temperature $\sim 1.5$~MK of ``hybrid plasma'', 
i.e. photo-ionized and collisionally excited) may have a comparable 
intensity to the \Kalpha resonance line \citep[see their Fig.11]{2000A&AS..143..495P}. 
Therefore, these two lines also may be excluded from further consideration, unless there is evidence of the presence 
of nearby, very high-density plasma of RX\, J0720.4$-$3125. 
 
From the measured value of  the equivalent width $\mathrm{W_E} = 1.35\pm0.3$~eV of the spectral 
line of \ion{O}{VII} we estimate  the number of ions (column density, N$_{\rm OVII}$) 
in the direction of RX\, J0720.4$-$3125  in two different ways. First, 
using the dependence of the equivalent width of the absorption feature on the
column density of \ion{O}{VII} \citep{2004ApJ...605..793F},  we estimated a lower limit 
of log~N$_{\ion{O}{VII}}=15.8-16.2~\mathrm{cm}^{-2}$ in the direction of RX\, J0720.4$-$3125. 

Second, we arrive at similar results,
log\, N$_{\mathrm{OVII}} \sim 15.9-16.1 \mathrm{cm}^{-2}$, using the general relationship
between measured equivalent width and number of atoms in the line of sight, 
$W_{\lambda}/\lambda=8.85 \times 10^{-13}~N_{j}\lambda f_{jk}$,
where f$_{jk}=0.695$ is the oscillator strength of the \ion{O}{VII} \Kalpha line 
(ATOMDB) and N$_{j}$ is the column density  \citep{1978ppim.book.....S}.

\begin{figure}
   \centering
   \includegraphics[width=7.0cm,clip=]{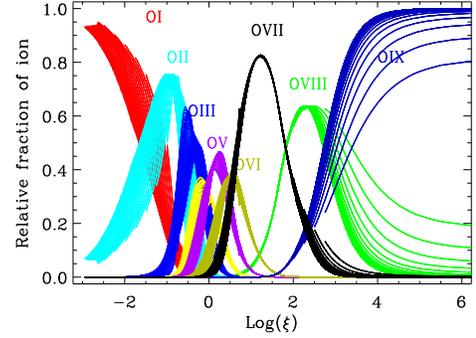}
      \caption{Ionization parameter  $\xi = \mathrm{L}_{X}/(\mathrm{n}_{H}r^2)$ of oxygen 
for different \nden{H}$=10^{1}-10^{9} \mathrm{cm}^{-3}$ volume densities located at the 
distance of $\mathrm{r}=10^{10} {\rm cm}$  from ionizing luminosity $\mathrm{L}_{X} \sim 10^{33} {\rm erg~s}^{-1}$ 
of a thermally emitting ($T \sim 10^6 \mathrm{K}$) INS.
              }
         \label{FigDzitta}
   \end{figure}

On the other hand, from the the average\footnote{Average n$_{H}$ hot gas density estimated 
in the direction of 7 low-mass X-ray binaries (\cite{2005ApJ...624..751Y}), 
assuming oxygen abundance relative to hydrogen $n(O)/n(H)=4~\times~10^{-4}$ \citep{1989GeCoA..53..197A} 
and an ionic fraction of \ion{O}{VII} $\sim$ 1 
(see, Fig.~\ref{FigDzitta}).} \nden{OVII}$=(1.35-2.84) \times 10^{-6} \mathrm{cm}^{-3}$ and the
distance to RX\, J0720.4$-$3125  \citep[$\mathrm{d} = 360^{+170}_{-90}$~pc,][]{2007ApJ...660.1428K} 
one obtains a value of 15.0$-$15.7 cm$^{-2}$ for log~N$_{\ion{O}{VII}}$. 
Moreover, using a very recent estimate of the spatial density of the hot gas on the Galactic 
plane ( \nden{H}$=(0.3-3.4) \times 10^{-3} \mathrm{cm}^{-3}$,\cite{2009ApJ...690..143Y}) 
based on {\emph Suzaku} observations, we have the value of 14.0$-$15.3 cm$^{-2}$ 
for log~N$_{Oxygen}$ in the direction of RX\, J0720.4$-$3125.

 This difference may be due to the origin of the absorption lines in the
immediate, ambient environment of the INS. If the absorption feature 
at 0.57~keV originates in the ionized plasma near the INS and belongs to highly ionized  \ion{O}{VII}, 
then the presence of the absorption line itself may constrain the possible location 
(separation from RX\, J0720.4$-$3125) of the ionized cloud. Indeed, according to
\cite{1969ApJ...156..943T}, if the plasma is located at a distance of $r$ 
from RX\, J0720.4$-$3125, the ionization parameter $\xi = L_{\rm X}/(n_{H}r^{2})=L_{\rm X}/(N_{H}r)$ 
is strongly determined by the  luminosity of $\sim 10^{33} {\mathrm{erg s}}^{-1}$. If so, 
oxygen will be fully photo-ionized at a distance $r < 10^{8} \mathrm{cm}$ \citep[][see, also Fig.~\ref{FigDzitta}]{1982ApJS...50..263K}.  

Moreover, in order to have a sufficient amount of \ion{O}{VII} ions near a thermally 
emitting ionizing source with temperature $\sim 10^6 \mathrm{K}$ and 
X-ray luminosity of $L_{\rm X}\sim10^{32}-10^{33} {\rm erg~s}^{-1}$, for the ionizing parameter the condition 
$1<\xi<10^3$ should be satisfied (Fig.~\ref{FigDzitta}). This means that a cloud/slab with rather 
high density (eg. \nden{H}$=10^{1}-10^{9} \mathrm{cm}^{-3}$) must be located within a distance of 
$10^{10}<r<10^{16} \mathrm{cm}$ from the ionizing source.

On the other hand, the hydrogen column density must be $< 2 \times 10^{20}$ ~cm$^{-2}$ (from the EPIC~pn spectroscopy),
ie. assuming oxygen abundance relative to the hydrogen $n(O)/n(H)=4~\times~10^{-4}$ \citep{1989GeCoA..53..197A}, 
we obtain for oxygen a total column density $N_{\rm Oxygen} < 8 \times 10^{16}$ ~cm$^{-2}$. 

Given the abovementioned general restrictions, a photo-ionized plasma associated with 
RX\, J0720.4$-$3125 may contribute to the observed fluxes in the UV and optical bands. 

The optical and ultraviolet fluxes observed from rather faint, optical counterparts of 
thermally emitting INSs, in general, lie a factor of $\sim$~10 above the extrapolated blackbody 
spectrum in X-rays \citep[e.g.][]{2008AIPC..983..331K}, the so-called ``optical excess''.
In the case of RX\, J1856.4$-$3754, the optical/UV spectrum is consistent with 
the slope of a Rayleigh-Jeans tail \citep{2001A&A...378..986V,2003A&A...399.1109B}, while for RX\, J0720.4$-$3125
there is a deviation and an additional power-law component was necessary to fit the data \citep{2003ApJ...590.1008K}.

In order to estimate the contribution from a photo-ionized plasma near an INS, as mentioned above,
we have used the widespread photoionization code CLOUDY  \citep[calculations were performed with 
version 07.02.02 of Cloudy, 
as described by][]{1998PASP..110..761F}. 
Despite the restrictions (luminosity of the ionizing source, ionization parameter, ie. a combination of 
location of the cloud relative to the source and spatial density of the medium) there are
still a number of free parameters for input to the CLOUDY code: 
Covering factor\footnote{The covering factor is the fraction 
of 4$\pi \, \mathrm{sr}$ covered by gas, as viewed from an ionizing source, ie. $\Omega/4\pi$. 
For details, see \cite{1998PASP..110..761F}, CLOUDY documents} 
(open or closed geometry), chemical abundances, filling factor, turbulence, etc. 

\begin{figure}
   \centering
   \includegraphics[angle=90,width=8.6cm,clip=]{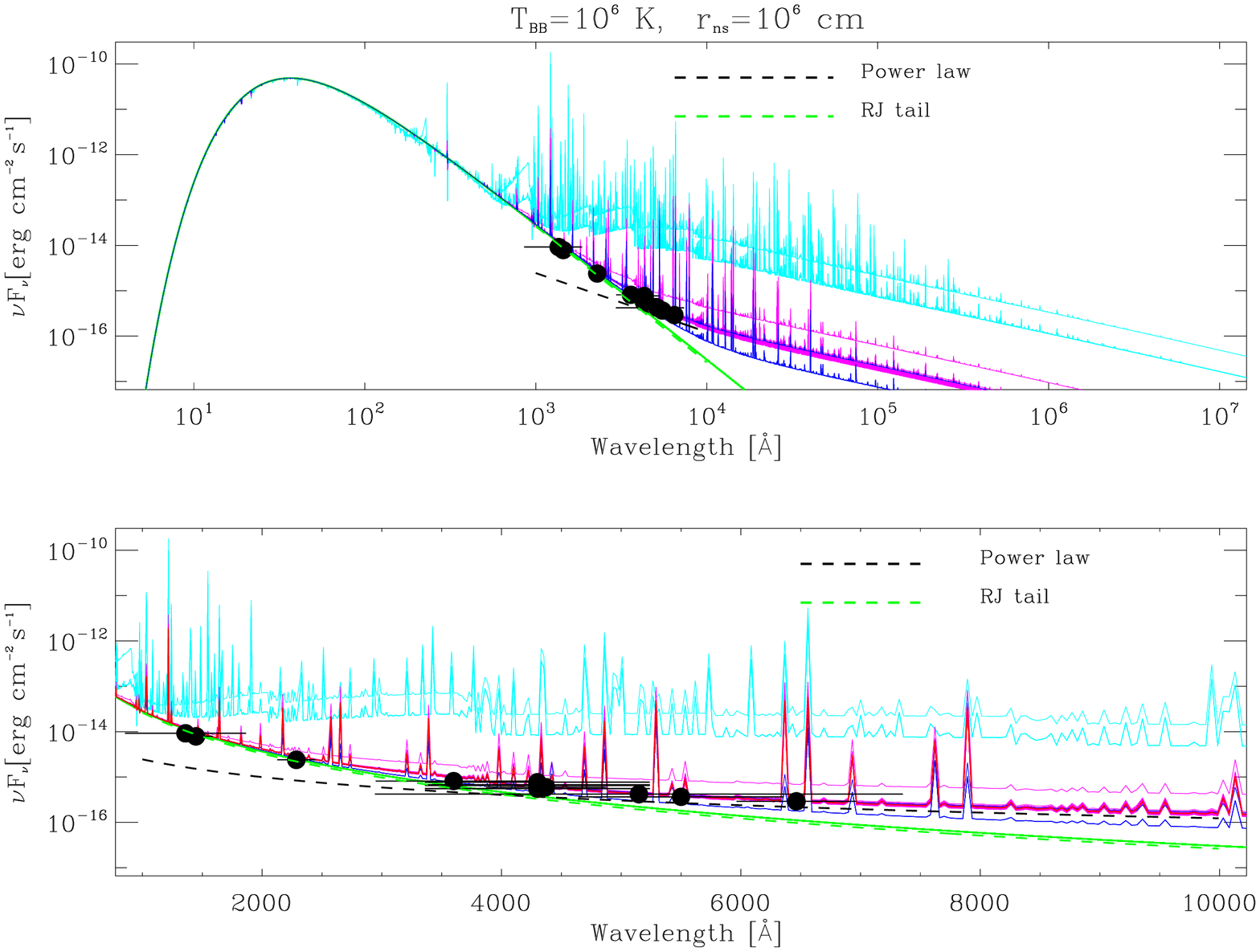}
   \caption{Simulated spectra of a photo-ionized plasma associated with a NS 
(blackbody spectrum  T=$10^{6}$ K and luminosity of $10^{33} {\rm erg~s}^{-1}$) for 
different initial hydrogen volume densities, located at the distance of $r_{inner}=10^{10}$cm.
The filled dots indicate the observed optical/UV fluxes of RX\, J0720.4$-$3125 \citep{1998ApJ...507L..49K,2003ApJ...590.1008K,2003A&A...408..323M,2008A&A_sub2}.
Fore some parameter sets the simulated spectra can explain the power-law component with an ``optical excess''.
              }
   \label{Figsim}
   \end{figure}

Nevertheless, as shown by our simulations, in principle, the existence of a 
small, relatively dense ambient medium
(e.g. \nden{H}$=3.5\times10^{8} \mathrm{cm}^{-3}$, covering factor $=\mathrm{0.013}$, located at 
the distance of $10^{10} \mathrm{cm}$ from INS, see Fig~\ref{Figsim}),
may at least partially explain the enigmatic ``optical excess'' observed at RX\, J0720.4$-$3125 
\citep{2003A&A...408..323M, 2003ApJ...590.1008K} and provide sufficient \ion{O}{VII} ions.

The satisfactory fit to the observed 
``optical excess'' was not unique and may also be provided with other sets of parameters.

A partial origin of the absorption feature in the local, interstellar hot medium 
(e.g. Local Bubble) cannot be excluded completely. However, 
 the {\em non-detection} of the absorption feature in other cases
\footnote{The absorption feature is absent in the co-added RGS spectrum 
of RX\ J1856.4-3754 ($\mathrm{d} = 161^{+18}_{-14}$~pc \citep{2007Ap&SS.308..191V}, 
but  clearly seen in the co-added spectra of RBS1223 and RX\, J1605.3+3249 with lower significance 
(see,  http://xmm.esac.esa.int/BiRD)} beyond the Local Bubble (for other objects, see, \citet{2005ApJ...624..751Y},  \citet{2006ApJ...644..174F}, \citet{2007ApJ...669..990B})
 makes the proposition of the origin of the absorption feature mainly in 
 the ambient medium of RX\, J0720-3125 more likely.

\section{Conclusions}
   \begin{enumerate}
      \item An absorption feature at 0.57~keV is clearly detected in the co-added {\em XMM-Newton} 
RGS spectrum of RX\, J0720.4$-$3125. The detected absorption feature likely is a blend.
      \item Most probably, it originates mainly in the ambient medium of RX\, J0720.4$-$3125 and 
may be identified with highly ionized oxygen (\ion{O}{VII}), 
consisting of Doppler shifted components. Neither an interstellar nor an atmospheric, partial origin can be excluded completely.
      \item The observed optical/UV flux excess of RX\, J0720.4$-$3125 compared to the extrapolated 
X-ray blackbody radiation partially may be caused by emergent emission of a nearby, relatively dense photo-ionized cloud. 

Further investigation, in particular  simultaneous X-ray and optical short-term variations could shed some light 
on the nature of this enigmatic object.
   \end{enumerate}

\begin{acknowledgements}
      VH and MMH acknowledge support by the German
      \emph{Deut\-sche For\-schungs\-ge\-mein\-schaft (DFG)\/} through project
      C7 of SFB/TR~7 ``Gravitationswellenastronomie''. We thank the anonymous referee for very important comments and discussion. 
\end{acknowledgements}

\bibliographystyle{aa} 
\bibliography{hambaryanetal.bib} 
\end{document}